\documentclass[12pt]{JHEP3}

\usepackage{amsmath,epsfig}
\usepackage{amssymb,amsfonts}
\usepackage{overpic}
\usepackage{xcolor}
\let\normalcolor\relax
\usepackage{graphicx}
\usepackage{longtable}
\usepackage{color}
\usepackage{bbm}
\usepackage{bm}
\usepackage{ulem}
\usepackage{array} 
\newcolumntype{C}{>{$}c<{$}}  

\usepackage{diagbox} 

\newcommand{\exclude}[1]{}
\def\nn{\nonumber}

\def\<{\langle}
\def\>{\rangle}

\def\d{\alphaathrm{d}}

\def\a#1{\alpha_{#1}}

\def\beq{\begin{equation}}
\def\eeq{\end{equation}}
\def\be{\begin{equation}}
\def\ee{\end{equation}}
\def\bea{\begin{eqnarray}}
\def\eea{\end{eqnarray}}
\def\bal{\begin{align}}
\def\eal{\end{align}}

\def\2b2[#1,#2][#3,#4]{\left( \begin{array}{cc} #1 & #2 \\ #3 & #4 \end{array}
\right)}
\def\3b3[#1,#2,#3][#4,#5,#6][#7,#8,#9]{\left( \begin{array}{ccc} #1 & #2 #3 \\
#4 & #5 & #6\\#7&#8&#9\end{array} \right)}

\newcommand{\g}{\gamma}

\newcommand\fverb{\setbox\pippobox=\hbox\bgroup\verb}
\newcommand\fverbdo{\egroup\alphaedskip\noindent%
                        \fbox{\unhbox\pippobox}\ }
\newcommand\fverbit{\egroup\item[\fbox{\unhbox\pippobox}]}

\newcommand{\bear}{\begin{eqnarray}}

\newcommand{\eear}{\end{eqnarray}}

\newcommand{\bsea}{\begin{subeqnarray}}
\newcommand{\esea}{\end{subeqnarray}}
\newbox\pippobox

\def\d{\delta}
\def\g{\gamma}

\def\6{\partial}
\def\a{\alpha}
\def\nn{\nonumber}

\def\e{\epsilon}
\def\m{\mu}
\def\n{\nu}
\def\r{\rho}
\def\s{\sigma}

\def\p{\partial}

\def\sq
\def\a{\alpha}
\def\b{\beta}

\def\e{\epsilon}

\def\d{\delta}

\usepackage{tensor}

\title{An Effective Theory Of Anomalous Momentum Diffusion From Holography}

\author{Jewel K.~Ghosh$^{\star \#}$, M. Arshad Momen$^{\star \# \dagger } $ \\
$^\star$ \href{http://www.iub.edu.bd/}{Independent University Bangladesh (IUB)}, Bashundhara RA, Dhaka 1229, Bangladesh \\
$^\#$ \href{https://ccds.ai/compas/} {Center for Computational and Data Sciences}, Independent University, Bangladesh, Bashundhara RA, Dhaka 1229, Bangladesh \\ 
$^\dagger$ \href{https://www.du.ac.bd/}{University of Dhaka}, Dhaka-1000, Bangladesh
 \\
Email: \href{mailto: jewel.ghosh@iub.edu.bd}{jewel.ghosh@iub.edu.bd}, \href{mailto:arshad@iub.edu.bd}{arshad@iub.edu.bd}, \href{mailto:amomen@du.ac.bd}{amomen@du.ac.bd}
}

\abstract{ We  consider a $U(1)$ Maxwell-Chern-Simons theory in $5$-dimensions, and analyze the vector perturbations around a classical charged black-brane background.  We solve the  equations of motion for these perturbations in a derivative expansion. By computing the boundary current, we find that time and spatial derivatives can be interpreted as the induced electric and magnetic field respectively, and the Chern-Simons term contributes to a nonzero divergence of the boundary current which indicates a quantum anomaly. Using holography, we  construct a two-derivative effective action for the vector perturbations. By complexifying the radial coordinate, and using appropriate transformation, we construct the full solution on the  complexified bulk contour. By computing the on-shell action for the full Schwinger-Keldysh geometry, we obtain the Keldysh functional. We find that the single boundary on-shell action mixes parity, whereas the Keldysh functional does not depend on the Chern-Simons term up to the quadratic orders in derivative expansion.   \\ \ \\ \ \\ \ \\ \ \\ \\ \ \\ \ \\ \ \\

}

\dedicated{Dedicated to Mrs. Shitali Ghosh, beloved mother of Jewel K. Ghosh.}
\begin{document}
\maketitle
\section{Introduction}
\begin{figure}[t]
\centering
\begin{overpic}[scale=.4]{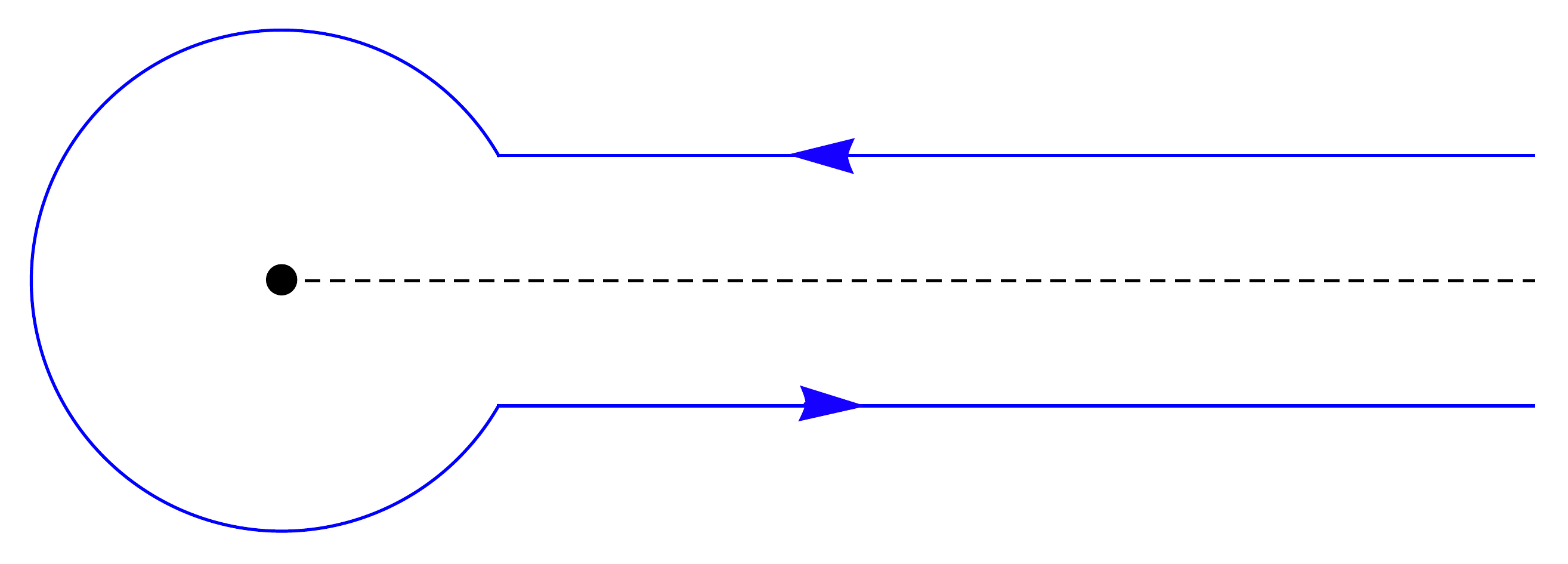}
\put(18,20){$r_+$}
\put(100,25){$\infty+i \epsilon$}
\put(100,9){$\infty-i \epsilon$}
\end{overpic}
\caption{Schwinger-Keldysh contour requires complexifying the bulk geometry.} \label{contour}
\end{figure}

The Anti-de Sitter (AdS)/Conformal Field Theory (CFT) correspondence \cite{Maldacena:1997re} (for review see \cite{Aharony:1999ti, Casalderrey-Solana:2011dxg}) relates a Quantum Field Theory in $d$-dimensions to a gravitational theory defined on higher dimensions. In the appropriate limit, the gravitational theory is approximated by classical General Relativity, where metric of the latter describes the ``bulk". Different fields in the bulk theory correspond to different operators of the  dual quantum field theory. The correspondence is further strengthened  by the equality of the partition functions as described by the GKPW rule \cite{Gubser:1998bc,Witten:1998qj}. \\
After the initial discovery, there have been several extensions in different directions. Bulks which are not AdS everywhere have been  studied. A particular class of example is asymptotically AdS space-time, as the name suggests the bulk asymptotes to AdS space-time near the boundary, though it can differ from AdS in the interior. It can have horizon, consequently, such black hole space-time carries a non-trivial Hawking temperature. On the dual side, the asymptotically AdS black hole corresponds to  an equilibrium quantum field theory at a finite temperature.  \\
To study non-equilibrium properties of quantum field theories, a natural tool is the Schwinger-Keldysh closed time path formalism \cite{calzetta_hu_2008, kamenev_2011}. Instead of defining the action on a positive real line for time, in the Schwinger-Keldysh formalism, the time runs forward and then backward in time forming a closed path. This is because of the fact that ket and bra in a density matrix evolves in opposite directions in time. In holography, the early attempts to capture the Schwinger-Keldysh physics was accomplished in \cite{Herzog:2002pc}, where the authors used Kruskal diagram to identify the field doubling as required by the Schwinger-Keldysh formalism.  Recently, a prescription \cite{Glorioso:2018mmw} has been proposed  to capture the Schwinger-Keldysh physics from the gravity dual. In this proposal the bulk geometry is complexified encircling the horizon. A cartoon of this geometry is shown in Fig. \ref{contour}.
Two segments of the bulk contour correspond to  the two different segments of the Schwinger-Keldysh closed path, and we specify the sources at the boundaries corresponding to each segment. Using this prescription, several studies have been conducted for different kinds of systems \cite{Chakrabarty:2019aeu, Loganayagam:2020eue, Jana:2020vyx, Loganayagam:2020iol, Ghosh:2020lel, He:2021jna, He:2022jnc, He:2022deg, Bu:2020jfo, Bu:2021jlp, Bu:2022esd, Banerjee:2022aub}. 
\\
Anomaly is an important concept  in quantum field theory which is defined to be any classical symmetry of the Lagrangian that is broken by quantum mechanical effects \cite{Bertlmann:1996xk, Fujikawa:2004cx}. Although the classical Lagrangian is invariant under a symmetry, the measure of the path integral may not be invariant of the symmetry under consideration, and hence the divergencelessness of the corresponding current does not hold quantum mechanically.  Different aspects of anomaly in holography have been extensively studied in the literature  \cite{Fukushima:2008xe, Kharzeev:2013ffa, Landsteiner:2016led, Kharzeev:2020jxw, Ghosh:2021naw}. \\
The purpose of this paper is to understand anomaly in Schwinger-Keldysh physics. As we have mentioned before, different bulk fields correspond to different boundary operators. In particular, a bulk gauge field described by a $U(1)$ Maxwell theory is dual to a boundary  current. Different aspects of the bulk gauge field have been extensively studied in the literature \cite{Policastro:2001yc, Policastro:2002se, Kovtun:2003wp}, and provided some early results emerging from holography. To describe anomaly in holography,  a standard approach is to add a Chern-Simons term. In total, this corresponds to the Maxwell-Chern-Simons theory for a $U(1)$ gauge field. Holographically, this gauge field is dual to an anomalous current on the boundary quantum field theory. We study  perturbation around the classical background, and then find an effective action for this perturbation. In the other way around, we derive a two-derivative action whose dynamics is dual to an anomalous current on the dual strongly coupled quantum field theory. After finding the solution of perturbation equation, we use an appropriate transformation to find the solution for the whole holographic Schwinger-Keldysh contour. By constructing the on-shell action, we find the Keldysh functional holographically. \\
The structure of this paper is as follows. In section \ref{setup}, we  describe the setup of the system that we study. We also describe some of the important properties of the background there. In section \ref{vecsec}, we analyze the perturbation around the background fields. The perturbation is classified as vector sector, and we show that it is the vector perturbation that gives rise to interesting dynamics and  affects the Chern-Simons term in the Lagrangian. We solve the perturbation equation in a derivative expansion, and the steps are spelled out in that section. We  also construct a $2$-derivative action variation of which  gives rise to the vector perturbation equations of motion. In section \ref{on-shell}, we compute the on-shell action, and renormalize it with appropriate counter-term. Section \ref{SK} is devoted for the Schwinger-Keldysh calculation. First we construct the full solution on the Schwinger-Keldysh radial contour, and then compute the Keldysh functional. 
\section{Setup} \label{setup}
In this work, we consider the Maxwell-Chern-Simons theory for a $U(1)$ gauge field in a $5$-dimensional space-time. The action we consider is \footnote{Our notation is the following: beginning small Latin letters (a,b,c...) will denote the bulk indices, Greek letters denote the boundary indices, middle small Latin  letters (i,j,m...) will denote the spatial indices. }$^,$\footnote{Convention for the Levi-Civita tensor is $\epsilon^{abcef}=-\frac{\varepsilon^{abcef}}{\sqrt{-g}}$, where $\varepsilon^{rtxyz}=1$. } 
\begin{equation}
S=-\int d^5x \sqrt{-g}\left( F_{ab}F^{ab}+\frac{\a}{4} \epsilon^{abcde}A_a F_{bc}F_{ef} \right). \label{action}
\end{equation}
Here 
\begin{equation}
A=A_a dx^a
\end{equation}
is the $U(1)$ gauge field and 
\begin{equation}
F=dA=\frac{1}{2}F_{ab}dx^a\wedge dx^b
\end{equation}
is the corresponding field-strength. The first part of the action \eqref{action} is the usual Maxwell part, whereas the second part is the Chern-Simons term with $\a$ being  the corresponding coupling. Note that the dimension of $\a$ is $-\frac{3}{2}$, and the theory is perturbatively non-renormalizable. \\
We are interested in studying an  anomalous current in a quantum field theory at a finite temperature. Holographically, the minimum ingredients we need a black-brane metric, and a non-trivial gauge field. To accomplish this, we take the following background \cite{Banerjee:2008th}
\begin{align}
& ds^2=-r^2 f(r) dv^2+2 dvdr+r^2 \d_{ij}dx^i dx^i, \label{metans} \\
& A=-\xi(r)dv,
\end{align}
where 
\begin{align}
& f(r)=1-\left(1+Q^2 \right) \left( \frac{r_+}{r} \right)^4 +Q^2 \left( \frac{r_+}{r} \right)^{6}, \label{metric}\\ & 
\xi(r)=\sqrt{\frac{3}{2}} Q \frac{r_+^{3}}{r^{2}}. \label{bg}
\end{align}
The metric is written in terms of ingoing Eddington-Finkelstein coordinates. The transformation that brings from the radial to Eddington-Finkelstein coordinates is described in Appendix \ref{Coord}. These field configurations satisfy the background equations: 
\begin{align}
& \nabla_b F^{ba}-\frac{3\a}{16}\epsilon^{abcde}F_{bc}F_{de}=0. 
\end{align}
The blackening factor $f(r)$ asymptotes to $1$ when $r\rightarrow \infty$. Near $r\rightarrow\infty$, the metric \eqref{metans} becomes: 
\begin{equation}
 ds^2\approx -r^2 dr^2+2dv dr+r^2 \d_{ij}dx^i dx^j
 \end{equation}
 which is the AdS$_{5}$ metric written in Eddington-Finkelstein coordinates. $r=\infty$ is the conformal boundary of the space-time, and near boundary, the metric asymptotes to the $5$-dimensional Anti-de Sitter space-time.  \\
 The metric \eqref{metric} describes a black-brane which is charged under the gauge field. By setting $Q=0$, we obtain the uncharged case, which is the pure black-brane metric. The blackening factor has a zero when $r=r_+$ which corresponds to a horizon. Temperature corresponding to this black-brane is given by
 \begin{equation}
   T=\frac{2-Q^2}{2\pi b}
   \end{equation}  
   where we have defined
   \begin{equation}
   b=\frac{1}{r_+}. 
   \end{equation}
Setting $Q=\sqrt{2}$, the Hawking temperature  vanishes. This emerges from the fact that, at this particular value of $Q$, there is a double zero of the blackening factor. As a consequence of the presence of a double zero, there will be an AdS$_2$ factor in the near horizon geometry. This corresponds to an extremal case. In general, we will call  \eqref{metans} to be Anti-de Sitter-Reissner-Nordstr\"om (AdS-RN) black-brane metric.  \\ 
 Equations \eqref{metric}-\eqref{bg} constitute the background for our system. In this work, we do not take gravity to be dynamical. That means we consider the metric to be the background, and any back-reaction on this will be ignored. In other words, we consider the probe limit, and the probe gauge field is described by the action \eqref{action}. Variation of the action gives
\begin{align}
\d S=& 4\int_{M} d^5x \sqrt{-g} \ \d A_a\left( \nabla_b F^{ba}-\frac{3\a}{16}\epsilon^{abcde}F_{bc}F_{de} \right) \nn\\
& -4 \int_{\partial M} d^4x\sqrt{-\g}\ n_a \d A_b  \left(F^{ab}+\frac{\a}{4} \epsilon^{abcde} A_c F_{de} \right) \label{firstvar}
\end{align}
where $M$ is the bulk manifold and $\partial M$ is its boundary with outward normal vector $n_a$. 
The gauge field satisfies the equation of motion
\begin{equation}
\nabla_b F^{ba}-\frac{3\a}{16}\epsilon^{abcde}F_{bc}F_{de}=0. \label{gfeom}
\end{equation}
 One can easily check that the background  \eqref{metric}-\eqref{bg} satisfy Eq. \eqref{gfeom}. In the following section we will study the fluctuation of the gauge field.  
\section{Vector perturbation} \label{vecsec}
In this section, we study the perturbation of the gauge field on top of the background given by Eq. \eqref{bg}. The perturbation can be classified as vector or scalar depending on their transformation under $SO(3)$ corresponding to the $3$-dimensional spatial part of the metric \eqref{metric}. Since we are considering a spin-1 field, the perturbation can be either vector or scalar. We start with the vector part, and comment on the scalar part in Appendix \ref{scalar}.  \\
In the vector sector, we perturb the gauge field\footnote{$\int_{\omega,k}$ is the shorthand notation for $\int \frac{d\omega}{2\pi} \frac{d^3k}{(2\pi)^3}$} 
 \begin{equation}
a_m= r \int_{\omega,k}  e^{-i \omega v} \left[ \Psi_{E} (r,k,\omega)  V_m^{E} (\vec{k}, \vec{x})+\Psi_{O} (r,k,\omega)  V_m^{O} (\vec{k}, \vec{x}) \right] . \label{vec} 
\end{equation}
where $V_m^{E,O}$ are the vector harmonics satisfying \cite{Gubser:2007nd, Kodama:2003kk}
\begin{equation}
\nabla^2 V_m^{E,O}+k^2 V_m^{E,O}=0, \ \ \p^m V_m^{E,O}=0. \label{defharmonic}
\end{equation}
Here $\nabla^2=\partial^m \partial_m$ is the $3$-dimensional Laplacian corresponding to the spatial part of the metric. The superscript refer to even (E) and odd (O) based on their behavior under parity transformations $\vec{x}\rightarrow -\vec{x}, \ \vec{k} \rightarrow -\vec{k}$. These vector harmonics have the following property: 
\begin{align}
& \p_m V_n^{E}-\p_n V_m^{E}=-i k \e_{mnp} V_p^{O}, \label{EO1} \\
& \p_m V_n^{O}-\p_n V_m^{O}=i k \e_{mnp} V_p^{E}. \label{EO2}
\end{align}
These equations are consistent in the sense that by using Eq. \eqref{defharmonic}, one equation maps the other. Note the sign difference on the right hand side of Eqs. \eqref{EO1}-\eqref{EO2}. This will manifest itself in the Chern-Simons contribution of the fluctuation equation which we describe below. \\
We use the following normalization condition
\begin{align}
& \int d^3 \vec{x}\  V^E_m \left( \vec{k}_1,\vec{x} \right) V_E^m \left( \vec{k}_2,\vec{x} \right)=\left(2\pi \right)^3 \delta^{(3)}\left( \vec{k}_1+\vec{k}_2\right), \\
& \int d^3 \vec{x}\  V^O_m \left( \vec{k}_1,\vec{x} \right) V_O^m \left( \vec{k}_2,\vec{x} \right)=-\left(2\pi \right)^3 \delta^{(3)}\left( \vec{k}_1+\vec{k}_2\right), \\
& \int d^3 \vec{x}\  V^O_m \left( \vec{k}_1,\vec{x} \right) V_E^m \left( \vec{k}_2,\vec{x} \right)=0.
\end{align}
This is consistent with Eqs. \eqref{EO1}-\eqref{EO2}.  Although we did not use any specific representation of the vector harmonics, a particular representation is given by \cite{Gubser:2007nd}
\begin{align}
& V^E_m \left(\vec{k},\vec{x} \right)=\frac{1}{k k_\perp}e^{i\vec{k}\cdot\vec{x}}\left(k_\perp^2\quad -k_xk_y\quad -k_x k_z \right) \\
& V^O_m \left(\vec{k},\vec{x} \right)= \frac{1}{ k_\perp}e^{i\vec{k}\cdot\vec{x}} \left(0\quad -k_z\quad k_y \right)
\end{align}
where 
\begin{equation}
\vec{k}=\left(k_x \quad k_y \quad k_z \right), \ \  k_\perp=\sqrt{k_y^2+k_z^2}, \ \ \text{and} \ \ k=\sqrt{k_x^2+k_y^2+k_z^2}. 
\end{equation}
Using this representation one can check all the statements made above. \\
Including the perturbation, in the vector sector, the gauge field and the field-strength tensors are: 
 \begin{align}
 & A=-\xi dv+\int_{\omega,k} \left(r\Psi_E V_j^E+r\Psi_O V_j^O \right)e^{-i \omega v}dx^j \\
 & F=-\xi' dr\wedge dv+\int_{\omega,k} \left[\left\{ \frac{d}{dr}\left(r\Psi_E \right)V_j^E+\frac{d}{dr}\left(r\Psi_O \right)V_j^O \right\} e^{-i \omega v} dr\wedge dx^j \right. \nonumber \\
 &\qquad\qquad\qquad\qquad\qquad -i \omega r \left(\Psi_E V_j^E+\Psi_O V_j^O \right)e^{-i \omega v} dv\wedge dx^j\nonumber \\
 & \qquad\qquad\qquad\qquad\qquad\left. -\frac{i kr}{2}\epsilon_{mnp} \left(\Psi_E V^O_p-\Psi_O V^E_p \right)e^{-i \omega v} dx^m\wedge dx^n\right] \label{F}
 \end{align}
  Using these, we can obtain the equation of motion for the vector fluctuation. They are: 
\begin{align}
& \frac{d}{dr}\left[ r^3 f \frac{d}{dr} \left( r \Psi_{E} \right) \right]=\left(k^2+3 i \omega r \right) \Psi_{E}+2 i \omega r^2 \frac{d \Psi_{E}}{dr}+\frac{3 i \a k \xi'r}{2} \Psi_{O}, \label{Vec1} \\
& \frac{d}{dr}\left[ r^3 f \frac{d}{dr} \left( r \Psi_{O} \right) \right]=\left(k^2+3 i \omega r \right) \Psi_{O}+2 i \omega r^2 \frac{d \Psi_{O}}{dr}-\frac{3 i \a k \xi'r}{2} \Psi_{E}. \label{Vec2}
\end{align}
A quick look at the Eqs. \eqref{Vec1} and \eqref{Vec2} shows that a complex combination of the
perturbations $\Psi_{E,O}$ will satisfy decoupled radial equations. These two equations got coupled because of the presence of the Chern-Simons term. As we mentioned before, the Chern-Simon term mixes the parity and effectively acts as a coupling between the even and odd sector. This is reflected in the equations \eqref{Vec1}-\eqref{Vec2}. Note also the sign difference in front of the Chern-Simons coupling in Eqs. \eqref{Vec1}-\eqref{Vec2}. This is because of Eqs. \eqref{EO1}-\eqref{EO2}, since they have opposite sign in the right hand side. 
\subsection{A two-derivative action for the vector perturbation}
In the previous section we had obtained the equation of motion for the vector fluctuations, which obeys a second order differential equation. It is therefore instructive to find a two-derivative action whose variation  will yield the fluctuation equation. \\
There are several ways to obtain this action. The straightforward procedure is to start with the action \eqref{action} and expand it upto quadratic order in the fluctuation $\Psi_{E,O}$. The quadratic part is the desired action. Performing this exercise we find: 
\begin{align}
& S^{(2)}= -2 \int_{\omega,k}\int dr \left[r^3 f\left\{  \frac{d}{dr}\left( r \Psi_E \right) \frac{d}{dr}\left( r \Psi_E^{(-)} \right)-\frac{d}{dr}\left( r \Psi_O \right) \frac{d}{dr}\left( r \Psi_O^{(-)} \right) \right\}  \right. \nn \\
&  +i \omega r^2 \left\{ \Psi_E^{(-)} \frac{d}{dr}\left( r \Psi_E \right) - \Psi_E \frac{d}{dr}\left( r \Psi_E^{{(-)}}\right) -  \Psi_O^{{(-)}} \frac{d}{dr}\left( r \Psi_O \right) + \Psi_O \frac{d}{dr}\left( r \Psi_O^{(-)} \right)\right\}\nn \\
& \left.  +k^2 r \left(\Psi_E \Psi_E^{(-)}-\Psi_O \Psi_O^{(-)} \right)\right] \nn \\
& +ik\a \int_{\omega,k}\int dr \left[ \frac{d}{dr} \left\{ r^2 \xi  \left(\Psi_E \Psi_O^{(-)}+\Psi_E^{(-)} \Psi_O \right) \right\}-3r^2\xi' \left(\Psi_E \Psi_O^{(-)}+\Psi_E^{(-)} \Psi_O \right) \right] \label{2dev}
\end{align}
where $\Psi^{{(-)}}_{E,O}\left(\omega,k \right)=\Psi_{E,O}\left(-\omega,k \right)|_{\vec{k}\rightarrow -\vec{k}} $.   
Note that, the $\alpha$-independent term arises from parity preserving Maxwell part, while the $\alpha$-dependent Chern-Simons term mixes parity. The total derivative part in the last line of Eq. \eqref{2dev} comes from the Chern-Simons part, and being  a total derivative term, this does not play a role in the equation of motion. It will, however, contribute to the on-shell action. We will elaborate more on this in section \ref{on-shell}. Starting from the action \eqref{2dev}, a straight-forward exercise reveals that the equation of motion for $\Psi_E$ and $\Psi_O$ are indeed Eqs. \eqref{Vec1}-\eqref{Vec2}. 
\subsection{Solution of perturbation equations in the hydrodynamic expansion}
In section \ref{vecsec}, we have found the equations for the vector fluctuation. Since the equations got coupled, this is unlikely that a closed form exact solution can be found. Instead, we work in the gradient expansion. Since the time and $\vec{x}$ dependence is exponential, a gradient expansion is also equivalent to small $\omega$ and $k$ expansion. This is what we refer to  the hydrodynamic expansion. \\
In the hydrodynamic expansion, we look for solution in the form
\begin{align}
& \Psi_E=\Psi_E^{(0)}+\left( \omega \Psi_E^{(\omega)}+k \Psi_E^{(k)}\right)+\left(\omega^2 \Psi_E^{(\omega^2)}+k^2 \Psi_E^{(k^2)}+\omega k \Psi_E^{(\omega k)}\right)+\cdots \label{expE} \\
& \Psi_O=\Psi_O^{(0)}+\left( \omega \Psi_O^{(\omega)}+k \Psi_O^{(k)} \right)+\left( \omega^2 \Psi_O^{(\omega^2)}+k^2 \Psi_O^{(k^2)}+\omega k \Psi_O^{(\omega k)}\right)+\cdots \label{expO}
\end{align}
where the superscript denotes the order of the perturbative expansion. For the pure Maxwell Lagrangian, no odd power of $k$ emerged. But due the presence of Chern-Simons term, odd power of $k$ can be non-trivial. This is a novel feature in the Chern-Simons case.  \\
We insert the expansions \eqref{expE} and \eqref{expO} into Eqs. \eqref{Vec1}-\eqref{Vec2}. Since the right hand side of  Eqs. \eqref{Vec1}-\eqref{Vec2} is higher order in derivative than the left hand side, equating at each order we have equations of the form: 
\begin{align}
& \frac{d}{dr}\left[ r^3 f \frac{d}{dr} \left( r \Psi_{E}^{\left( \omega^m k^n\right) 
} \right) \right]=J_{E}^{\left( \omega^m k^n \right)
} (r), \label{GenE} \\
& \frac{d}{dr}\left[ r^3 f \frac{d}{dr} \left( r \Psi_{O}^{\left( \omega^m k^n \right)
} \right) \right]=J_{O}^{\left( \omega^m k^n \right)}(r) \label{GenO}
\end{align}
where the right hand side represents the source that is built from the solutions of previous orders. The only exception is the zeroth order for which the equation is sourceless, i.e the right hand side is zero.  \\
Let us now take the equations at each order, in which the zeroth order equation is: 
\begin{equation}
\frac{d}{dr}\left[ r^3 f \frac{d}{dr} \left( r \Psi_{E,O}^{\left( 0 \right) 
} \right) \right]=0.
\end{equation}
This is a second order differential equation, integration of which contains two integration constants.  As we want to impose regularity at the horizon $r=r_+$, one integration constant has to be set to zero. Here regularity guarantees the analyticity of the quantity at the horizon. Therefore, the solution to the zeroth order equation is 
\begin{equation}
 \Psi_{E,O}^{\left( 0 \right) }=\frac{C_{E,O}}{r}. 
\end{equation}
Following the usual holographic dictionary, $C_{E,O}$ will be identified as the source on the boundary. \\
For non-zero order, we can write a general expression  for the solution of equations Eqs. \eqref{GenE}-\eqref{GenO}. Again, these are second order differential equations, which require two boundary conditions. We demand regularity at the horizon. We also require that $r\Psi_{E,O}$ goes to $C_{E,O}$ at the boundary. Since we have already demanded that $r \Psi_{E,O}^{(0)} $ goes to $C_{E,O}$ at the boundary, this implies that $ r \Psi_{E,O}^{(\omega^m k^n)} $ where both $m,n\neq 0$, goes to zero at the boundary. After imposing these two conditions,  some manipulations yield the following general expression for the solution: 
\begin{align}
\Psi_{E,O}^{\left( \omega^m k^n \right)}&=-b^2 \frac{1}{br}\int_{br}^{\infty}d\eta \ \frac{1}{\eta^3 f\left(\frac{\eta}{b} \right)} \int_1^\eta d\kappa\  J_{(E,O)}^{\omega^m k^n 
}\left( \frac{\kappa}{b} \right) \label{gen}  \\
&=-b^2 \frac{1}{br}\int_{br}^{\infty}d\eta \ \frac{1}{\eta^3 f\left(\frac{\eta}{b} \right)} \left\{ F^{\left( \omega^m k^n\right) }(\eta)-F^{\left( \omega^m k^n\right)}(1) \right\}  \\
& =b^2 \frac{1}{br}\int_{br}^{\infty}d\eta \ \frac{F^{\left(\omega^m k^n\right)}(1)}{\eta^3 f\left(\frac{\eta}{b} \right)} -b^2 \frac{1}{br}\int_{br}^{\infty}d\eta \ \frac{F^{\left(\omega^m k^n\right) }(\eta)}{\eta^3 f\left(\frac{\eta}{b} \right)}, \label{gensim}
\end{align}
where we have defined
\begin{equation}
F^{\left(\omega^m k^n \right)} (\eta)=\int_1^\eta d\kappa\  J_{(E,O)}^{\omega^m k^n 
}\left( \frac{\kappa}{b} \right). 
\end{equation}
The first term in Eq. \eqref{gensim} is universal, and can be integrated to find: 
\begin{align}
\int_{br}^\infty d\eta \frac{1}{\eta^3 f\left(\frac{\eta}{b} \right)}=&-\frac{i \pi  \left(-2 Q^2+2 Z+1\right)}{4 \left(Q^2-2\right) Z}-\frac{\left(1-2 Q^2\right) \tanh ^{-1}\left(\frac{2 b^2 r^2+1}{Z}\right)}{2  \left(Q^2-2\right) Z} \nn \\ 
& -\frac{\log \left(b^4 r^4+b^2 r^2-Q^2\right)-2 \log \left(1-b^2 r^2\right)}{4 \left(Q^2-2\right)}
\end{align}
where we have defined $Z=\sqrt{1+4Q^2}$. Near $r=\infty$, this has an asymptotic expansion:
\begin{equation}
\int_{br}^\infty d\eta \frac{1}{\eta^3 f\left(\frac{\eta}{b} \right)}\underset{r\rightarrow\infty}{=}\frac{1}{2 (b r)^2}+\frac{1+Q^2}{6 (b r)^6}+\mathcal{O}\left(\frac{1}{(br)^7} \right).
\end{equation}
The main difficulty to find an exact solution comes from the second part of Eq. \eqref{gensim}. However, often we need an expansion near the boundary located at $r=\infty$. We circumvent the difficulty by expanding the integrand of the second term in Eq. \eqref{gensim} near $r=\infty$, and then perform the integration. One might  worry that the constant term might be different in this way of finding the near boundary expansion. This is fixed since we set the constant term for non-zero order to be zero, which is manifest from the upper limit of Eq. \eqref{gen}.  Now we perform the integrations when possible to find the solution, and then find the near boundary expansion. In the case where the integrations are not tractable, we perform the strategy outline above to obtain the near boundary expansion. \\
The details of the calculations are presented in Appendix \ref{details}. Here, we record the main results. For the linear order, we could find the exact solutions. They are: 
\begin{align}
& \Psi^{(\omega)}_{E,O}(r)=-i b^2 \frac{C_{E,O} }{br} F_1(br), \label{solomega} \\
& \Psi^{(k)}_{E,O}(r)=\pm \left( \frac{3}{2} \right)^{3/2}i \a C_{O,E} Q b^2 \frac{1}{br} F_2 (br), \label{solk}
\end{align} 
where we have defined
\begin{align}
& F_1(br)=\int_{br}^\infty d\eta \frac{\eta-1}{\eta^3 f\left(\frac{\eta}{b} \right)}, \\
& F_2 (br)=\int_{br}^{\infty} d\eta \frac{\eta^2 -1}{\eta^5 f\left( \frac{\eta}{b} \right)}.
\end{align}
Note that Eq. \eqref{solk} originates from the Chern-Simons terms. In the absence of Chern-Simons term or for uncharged solution, this term would not be present. \\
The integrations becomes intractable for quadratic terms. However, by the strategy outlined above we have obtained the near boundary $r\rightarrow\infty$ expansion. They are: 
\begin{align}
& \Psi_{E,O}^{(\omega^2)}=\frac{b^3}{4} C_{E,O}\left\{\frac{4 G_1-1}{(br)^3}+\frac{2 \ln (br)}{(br)^3} \right\}+\frac{ b^3}{2}C_{E,O} \frac{1}{(br)^4}+\mathcal{O}\left(\frac{1}{r^7} \right), \\
& \Psi_{E,O}^{(k^2)}= -b^3 C_{E,O} \frac{27 \alpha ^2 \ln 2+16 \ln (b r)-16 G_2+8}{32 (br)^3} +\mathcal{O}\left(\frac{1}{r^7} \right) \\
& \Psi_{E,O}^{(\omega k)}=\pm b^3 \left( \frac{3}{2}\right)^{\frac{1}{2}} \alpha C_{O,E} \left\{ \frac{f_3 (Q)}{(br)^3}+\frac{3Q}{4 (br)^4}-\frac{3Q}{8 (br)^6} \right\}+\mathcal{O}\left(\frac{1}{r^7} \right)
\end{align}
where $G_1$ and $G_2$ are the values of integrals $\int d\kappa \sqrt{\kappa}\frac{d}{d\kappa}\left\{ \sqrt{\kappa} F_1(\kappa) \right\}  $ and $G_2$ is the value of $\int d\kappa \left\{\frac{1}{\kappa}-\frac{27\a^2 Q^2}{4 \kappa^3}F_2 (\kappa) \right\}$ respectively evaluated at $\kappa=1$ which is the horizon. The function $f_3(Q)$ is defined in Eq. \eqref{f3}. This vanishes at $Q=0$. \\
Now we have all the ingredients to write the near boundary expansion of the fluctuation. We find
\begin{align}
\frac{\Psi_{E,O}}{b}=&  C_{E,O }\frac{1}{b r}-ib \omega C_{E,O}  \left\{ \frac{1}{(br)^2}-\frac{1}{2 (br)^3}+\frac{1+Q^2}{5 (br)^6}+\mathcal{O}\left(\frac{1}{(br)^7} \right) \right\} \nn \\
& \pm i b k \left( \frac{3}{2} \right)^{3/2}  \a C_{O,E} Q \left\{ \frac{1}{(br)^2}-\frac{1}{2 (br)^3}+\frac{1+Q^2}{5 (br)^6} + \mathcal{O}\left(\frac{1}{(br)^7} \right)  \right\} \nn \\
& +\frac{b^2 \omega^2}{4} C_{E,O} \left\{  \left(  \frac{4 G_1-1}{(br)^3}+\frac{2 \ln (br)}{(br)^3} \right)+ \frac{1}{2 (br)^4}+\mathcal{O}\left(\frac{1}{(br)^7} \right) \right\} \nn \\
& -b^2 k^2 C_{E,O} \left\{ \frac{27 \alpha ^2 \ln 2+16 \ln (b r)-16 G_2+8}{32 (br)^3} +\mathcal{O}\left(\frac{1}{(br)^7} \right) \right\} \nn \\
& \pm b^2 \omega k  \left( \frac{3}{2}\right)^{\frac{1}{2}} \alpha C_{O,E} \left\{ \frac{f_3 (Q)}{(br)^3}+\frac{3Q}{4 (br)^4}-\frac{3Q}{8 (br)^6} +\mathcal{O}\left(\frac{1}{(br)^7}  \right) \right\}\nn\\
& \qquad\qquad\qquad\qquad\qquad\qquad\qquad\qquad \qquad+\mathcal{O}\left(\omega^3,k^3, \omega k^2, k^2\omega \right).
\end{align}
We will require this expression for various holographic calculations. In the next section, we compute the on-shell action, and then renormalize it. 
\section{Quadratic on-shell action}
 \label{on-shell}
 In this section, we will obtain the on-shell action. Following the AdS/CFT dictionary, this on-shell action corresponds to the generating functional of the dual quantum field theory. We start with the bare action, and then proceed to renormalize by using the holographic renormalization. 
\subsection{Bare action}
The bare action can be obtained in a couple of ways. One can start from the first variation of the action recorded in Eq. \eqref{firstvar}, and impose the on-shell condition. The latter kills the first line of Eq. \eqref{firstvar} and we find: 
\begin{equation}
 S^{(1)}_{\text{on-shell}}=-4 \int_{\partial M} d^4x\sqrt{-\g}\ n_a \d A_b  \left(F^{ab}+\frac{\a}{4} \epsilon^{abcde} A_c F_{de} \right). \label{firstvaronshell} 
 \end{equation} 
 Here $n_a$ is the outward unit normal to the regulated boundary defined via $r=\Lambda$, where $\Lambda\rightarrow \infty$ for the boundary, and $\g_{\m\n}$ is the induced metric on the boundary. A straight-forward calculation gives
 \begin{equation}
 n_a=\left(\frac{1}{r\sqrt{f}},0,\vec{0} \right). 
 \end{equation}
 Since the variation of the gauge field does not change the shape of the boundary, the normal vector is also unchanged. A variation of Eq. \eqref{firstvaronshell} gives
 \begin{equation}
 S^{(2)}_{\text{on-shell}}=-2 \int_{\partial M}d^4x  \sqrt{-\g} n_a \d A_b \left\{ \d F^{ab}+\frac{\a}{4} \epsilon^{abcde}\left(\d A_c \bar{F}_{de}+\bar{A}_c \d F_{de} \right)\right\}
 \end{equation}
 where a over-bar denotes the background (unperturbed) value. \\
 Alternatively, one can start with the action \eqref{2dev}. There will be some total derivative terms which become the boundary terms. In either way we find a simple result
 \begin{align}
 S^{(2)}_{\text{on-shell}}=& -2 \int_{\omega,k} r^4 f \left\{ \Psi_E \frac{d}{dr}\left( r\Psi_E^{(-)} \right)-\Psi_O \frac{d}{dr}\left( r\Psi_O^{(-)} \right) \right\} \nn \\
 & +i \a r^2 \xi \int_{\omega,k} k \left( \Psi_E \Psi_O^{(-)}+\Psi_E^{(-)} \Psi_O\right). \label{S2}
 \end{align}
 The first line comes from the Maxwell part, and the second line is the Chern-Simons contribution. Once again, the Maxwell part is parity preserving whereas the Chern-Simons part
mixes parity. As this is a quadratic action for the perturbations, a suitable diagonalied modes involving the
mixing of the modes $\Psi_{E,O}$ will produce decoupled equation as we had commented earlier. This action is the generating functional of the dual QFT via the AdS/CFT correspondence. 
\subsection{Holographic renormalization}
The on-shell action is divergent. To remove the divergences we need appropriate counter-term. As described in \cite{Landsteiner:2011iq}, the required counter-term is: 
\begin{equation}
S_{\text{ct}}= u \int d^4x \sqrt{-\g}\hat{F}_{(0)\m\n}\hat{F}^{(0)\m\n} \label{ct}
\end{equation}
where the bulk metric is written in domain wall coordinates as: 
\begin{equation}
ds^2=du^2+\g_{\m\n}dx^\m dx^\n. 
\end{equation}
In Eq. \eqref{ct}, $\hat{F}_{(0)\m\n}$ is the leading term in the near boundary expansion. Now we need a relationship between the coordinates $r$ and $u$ near the boundary. \\
Following Appendix \ref{Coord}, a coordinate transformation
 \begin{equation}
 du=\frac{dr}{r \sqrt{f(r)}}
 \end{equation}
takes the metric from the Eddington-Finkelstein coordinates to the domain wall coordinates. Near the boundary $f(r)\rightarrow 1$, therefore the required coordinate transformation is: 
\begin{equation}
u=\ln (mr)+\mathcal{O} \left(\frac{1}{r} \right). 
\end{equation}
where $m$ is introduced for dimensional reason. Since the only dimension we have is the temperature, we set the scheme by choosing $m=b$. Therefore, the required counter-term is: 
\begin{equation}
S_{\text{ct}}= \ln (br) \int d^4x \sqrt{-\g}\hat{F}_{(0)\m\n}\hat{F}^{(0)\m\n}. \label{ctneeded}
\end{equation}
Since this action is quadratic in the field-strength tensor, the renormalization is required starting from the second order in fluctuation. \\
We can extract $\hat{F}_{\m\n}$ from Eq. \eqref{F}.  Explicitly, 
\begin{align}
\hat{F}=-\int_{\omega,k} \left[\frac{}{} i \omega r \left(\Psi_E V_j^E+\Psi_O V_j^O \right)e^{-i \omega v} dv\wedge dx^j \right. \nn \\
\left. +\frac{i kr}{2}\epsilon_{mnp} \left(\Psi_E V^O_p-\Psi_O V^E_p \right)e^{-i \omega v} dx^m\wedge dx^n \right]. \label{Fhat}
\end{align}
From here we can extract: 
\begin{align}
& \hat{F}_{vj}=-\int_{\omega,k} \frac{}{} i \omega r \left(\Psi_E V_j^E+\Psi_O V_j^O \right)e^{-i \omega v}, \\
& \hat{F}_{mn}=-\int_{\omega,k}i kr\epsilon_{mnp} \left(\Psi_E V^O_p-\Psi_O V^E_p \right)e^{-i \omega v}.
\end{align}
Using these, a straightforward calculation yields
\begin{equation}
S_{\text{ct}}=-2  r^2\ln (br) \int_{\omega,k} \left(\frac{\omega^2}{\sqrt{f}}-k^2 \sqrt{f}\right) \left(\Psi_E \Psi_E^{(-)}-\Psi_O  \Psi_O^{(-)} \right). \label{Sct}
\end{equation}
Note the the counter-term starts from the quadratic order in the hydrodynamic expansion. 
\subsection{Renormalized action}
Now we have all the ingredients to compute the renormalized on-shell action.  The latter is given by: 
\begin{equation}
S^{\text{ren}}_{\text{on-shell}}=S^{(2)}_{\text{on-shell}}+S_{\text{ct}}.
\end{equation}
Using Eqs. \eqref{S2} and \eqref{Sct} we find the following expression for the renormalized on-shell action
\begin{align}
S^{\text{ren}}_{\text{on-shell}}=& \int_{\omega,k} \left[-2 r^4 f \left\{ \Psi_E \frac{d}{dr}\left( r\Psi_E^{(-)} \right)-\Psi_O \frac{d}{dr}\left( r\Psi_O^{(-)} \right) \right\}+i \a k r^2 \xi \left( \Psi_E \Psi_O^{(-)}+\Psi_E^{(-)} \Psi_O\right) \right. \nn \\
&\left. -2 r^2 \ln (br) \left(\frac{\omega^2}{\sqrt{f}}-k^2 \sqrt{f}\right) \left(\Psi_E \Psi_E^{(-)}-\Psi_O  \Psi_O^{(-)} \right)\right]. \label{Srenfin}
\end{align}
This will be the generating functional on the QFT side via the AdS/CFT correspondence. We can observe that both the Maxwell and the counter-term parts of the renormalized on-shell action \eqref{Srenfin} are parity preserving, but the Chern-Simons part mixes parity. \\
We insert the near boundary expansion of $\Psi_{E,O}$ into Eq. \eqref{Srenfin}. Since $\xi\sim \frac{1}{r^2}$ and $\Psi_{E,O} \sim \left\{ \frac{1}{r}+\mathcal{O}\left( \frac{1}{r^2}\right) \right\}$, the Chern-Simons part does not contribute on-shell action directly. It does, however, through the solution. Therefore, all the divergences come from the Maxwell term, which get canceled by the counter-term. After the cancellation we get: 
\begin{align}
S_{\text{on-shell}}^{\text{ren}}= \frac{3\sqrt{6} i C_E C_O Q \a}{b}k+\frac{C_E^2-C_O^2}{8} \left\{ 32 G_1 \omega^2 +\left(16 G_2-27\a^2 \ln 2 \right)k^2 \right\} \nn \\
+\mathcal{O}\left(\omega^3, \omega^2 k, \omega k^2, k^3 \right), \label{Srenonshell}
\end{align}
where $G_1$ and $G_2$ are defined in Eqs. \eqref{G1} and \eqref{G2} respectively. In Eq. \eqref{Srenonshell} all the terms proportional to $\a$ are coming from the Chern-Simons term. Once again, the parity mixing term begets from the Chern-Simons term. Consequently, the even-odd correlator is non-trivial in this case. 
\subsection{The boundary current}   
To get more insight about the derivative expansion, in this section, we will compute the boundary current. According to the standard AdS/CFT prescription, the boundary current is given by \cite{Zaanen:2015oix, Ammon:2015wua}: 
\begin{equation}
\langle J^\m \rangle=\sqrt{-\gamma} n_r \left( F^{r\m}+\frac{\a}{4}\varepsilon^{r\m\n\r\s}A_\n F_{\r\s} \right).
\end{equation}
Using the results from previous section, this can be written as
\begin{equation}
\langle J^\m \rangle=r^3 \left( F^{r\m}+\frac{\a}{4}\varepsilon^{r\m\n\r\s}A_\n F_{\r\s} \right).
\end{equation}
Using different expressions and after some algebra we find: 
\begin{align}
& \langle J^v \rangle=r^3 \left[\xi'+\frac{i\a}{2r}\int_{\omega_1,\vec{k}_1} \int_{\omega_2,\vec{k}_2}  k_2 \left\{ \Psi_{O}^{(1)} \Psi_{E}^{(2)} \vec{V}_{O}^{(1)}\cdot \vec{V}_{O}^{(2)}-\Psi_{E}^{(1)} \Psi_{O}^{(2)} \vec{V}_{E}^{(1)}\cdot \vec{V}_{E}^{(2)}  \right. \right. \nn \\
&\qquad\qquad\qquad \left. \left. + \Psi_E^{(1)}\Psi_E^{(2)} \vec{V}_E^{(1)}\cdot \vec{V}_O^{(2)}-\Psi_O^{(1)}\Psi_O^{(2)} \vec{V}_O^{(1)}\cdot \vec{V}_E^{(2)}  \right\} e^{-i(\omega_1+\omega_2)v} \right] \\
& \langle J^m \rangle=r^3 \left[\int_{\omega,k} \left\{ \left( f \frac{d}{dr} \left( r\Psi_E \right)-\frac{i\omega}{r} \Psi_E\right)V^m_E+ \left( f \frac{d}{dr} \left( r\Psi_O \right)-\frac{i\omega}{r} \Psi_O\right)V^m_O \right\} e^{-i\omega v} \right. \nn \\
&\qquad\qquad -\frac{\alpha}{4 r^3} \left\{-2ir\xi \int_{\omega,k} k \left( \Psi_E V^m_O-\Psi_O V^m_E \right) e^{-i\omega v} \right.  \nn \\
&\qquad\qquad -2ir^2 \int_{\omega_1, k_1} \int_{\omega_2, k_2} \epsilon^{mnp} \omega_1 \left(\Psi_E^{(1)}\Psi_E^{(2)} V_n ^{E(1)} V_p ^{E(2)}  +\Psi_O^{(1)}\Psi_O^{(2)} V_n ^{O(1)} V_p ^{O(2)} \right. \nn \\
&
\qquad\qquad \left. \left. \left. +\Psi_E^{(1)}\Psi_O^{(2)} V_n ^{E(1)} V_p ^{O(2)}+\Psi_O^{(1)}\Psi_E^{(2)} V_n ^{O(1)} V_p ^{E(2)} \right)e^{-i(\omega_1+\omega_2)v} \right\} \right].
\end{align}
To see the implications of the hydrodynamic expansion, we compute the divergence of the boundary current. After some algebra, we find:
\begin{align}
\langle \partial_\m  J^\m \rangle&=2 \a r^2 \int_{\omega_1,k_1} \int_{\omega_2,k_2} \omega_1 k_2 \left\{  \Psi_{O}^{(1)} \Psi_{E}^{(2)} \vec{V}_{O}^{(1)}\cdot \vec{V}_{O}^{(2)}-\Psi_{E}^{(1)} \Psi_{O}^{(2)} \vec{V}_{E}^{(1)}\cdot \vec{V}_{E}^{(2)} \right. \nn \\
& \left. + \Psi_E^{(1)}\Psi_E^{(2)} \vec{V}_E^{(1)}\cdot \vec{V}_O^{(2)}-\Psi_O^{(1)}\Psi_O^{(2)} \vec{V}_O ^{(1)}\cdot \vec{V}_E^{(2)}  \right\} e^{-i(\omega_1+\omega_2)v}. \label{ano}
\end{align}
This can be succinctly written 
\begin{align}
\langle\partial_\mu J^\mu \rangle=-\frac{\a}{4} \epsilon^{\m\n\r\s}\hat{F}_{\m\n}\hat{F}_{\r\s}=-\a \hat{F}\wedge \hat{F}. \label{anoFwF}
\end{align} 
That means inserting expression of $\hat{F}$ from Eq. \eqref{Fhat} into Eq. \eqref{anoFwF}, we can retrieve Eq. \eqref{ano}.   This is the expression of familiar ABJ anomaly in Quantum Field Theory \cite{Adler:1969gk, Bell:1969ts, Peskin:1995ev}. Note that there was no electric field or magnetic field in the background. It's the fluctuation that induces an effective electric $(\sim \omega)$ and magnetic field $(\sim k)$. Therefore, we can interpret the current $J^\m$ to be anomalous where the symmetry is broken by the gradient expansion. The Chern-Simons term in the action takes care of this symmetry breaking.  
\section{Schwinger-Keldysh contour and the influence functional} \label{SK}
In the previous section we have used the ingoing Eddington-Finkelstein coordinates. To construct the full Schwinger-Keldysh solution we should also patch the the outgoing Hawking solution. The procedure to obtain the outgoing solution from the ingoing solution is presented below \cite{Jana:2020vyx}. \\
 The metric we started with is: 
 \begin{equation}
ds^2=2dvdr-r^2 f(r)dv^2+r^2 dx_i dx^i \ . \label{metricin}
\end{equation}
This was written in ingoing Eddington-Finkelstein coordinates. We define a new coordinate $\chi$ by the following equation: 
\begin{equation}
dr=\frac{i\b}{2} r^2 f(r) d\chi \label{chi}
\end{equation}
where 
\begin{equation}
\b=\frac{2\pi b}{\left[2-Q^2 \right]}
\end{equation}
is the inverse Hawking temperature. The metric \eqref{metricin} can be written as: 
\begin{equation}
ds^2=\frac{i\b}{2} \left(r^2 f \right)^2 dv \left( i\b d\chi-dv\right)+r^2 dx_i dx^i.
\end{equation}
From here this is manifest that: 
 \begin{equation}
dv\to  i\b d\chi-dv=dv', \ \omega\to -\omega \ .  
\end{equation}
preserves the form of the metric. This coordinate transformation will be referred as the Debye time reversal, and it takes from the ingoing to the outgoing Eddington-Finkelstein coordinates. Under this time reversal the vector perturbations do not change since they only concern the spatial coordinates. Therefore the full solution is given by: 
\begin{align}
& \Psi_E^{SK}=\Psi_E^{in}|_{\left\{ C_E\rightarrow C_{\text{in}}^E, C_O\rightarrow C_{\text{in}}^O \right\} }+ e^{-\omega\b\chi} \Psi_E^{out}|_{\left\{ C_E\rightarrow C_{\text{out}}^E, C_O\rightarrow C_{\text{out}}^O \right\} } , \\
& \Psi_O^{SK}=\Psi_O^{in}|_{\left\{ C_E\rightarrow C_{\text{in}}^E, C_O\rightarrow C_{\text{in}}^O \right\} }+ e^{-\omega\b\chi} \Psi_O^{out}|_{\left\{ C_E\rightarrow C_{\text{out}}^E, C_O\rightarrow C_{\text{out}}^O \right\} }.
\end{align}
Here $\Psi^{in}_{E,O}$ are the solutions that we have constructed so far in the ingoing Eddington-Finkelstein coordinates. $\Psi^{out}_{E,O}$ is constructed from $\Psi^{in}_{E,O}$ by taking $\omega\rightarrow -\omega$.  We can also write the coefficients $C_{in}^{E,O}, C_{out}^{E,O}$ in terms of the Schwinger-Keldysh basis $C_L^{E,O}, C_R^{E,O}$. Holographically this amounts to complexifying the bulk geometry as Fig. \ref{contour} shows. 
From Eq. \eqref{chi}, by a straightforward complex integration by using Cauchy integral formula we can write: 
\begin{equation}
\chi(\infty+i \epsilon)=0, \quad \chi(\infty-i \epsilon)=1. 
\end{equation}
The loci $\chi=0$ and $\chi=1$ will be denoted as the left and right boundaries respectively. We can specify the sources at the left and right boundaries which we denote as the left/right basis. In the left/right basis we have the following equations:
\begin{align}
& C_{in}^{E,O}+C_{out}^{E,O}=C_L^{E,O}, \\
& C_{in}^{E,O}+C_{out}^{E,O}e^{-\omega\b}=C_R^{E,O}. \label{LR}
\end{align}  
Solution of these is given by: 
\begin{align}
C_{in}^{E,O}=f_{BE}(\omega) \left(C_R^{E,O} e^{\b\omega}-C_L^{E,O} \right), \quad C_{out}^{E,O}=f_{BE}(\omega)\left(C_L^{E,O}-C_R^{E,O} \right)e^{\b\omega},  
\end{align}
where 
\begin{equation}
f_{BE}(\omega)=\frac{1}{e^{\b\omega}-1}
\end{equation}
is the Bose-Einstein factor. \\
We will now proceed to calculate the on-shell action \eqref{Srenfin} for the full Schwinger-Keldysh solution. For the complexified geometry, there are two boundaries located at $\chi=0$ and $\chi=1$. To get the full on-shell action we consider the individual on-shell actions, and subtract. We subtract because the unit normals are directed opposite at the boundaries.  That means, the Schwinger-Keldysh functional is given by: 
\begin{equation}
S_{\text{on-shell}}^{\text{SK}}=S_{\text{on-shell}}^{\text{ren}}|_{\chi=1}-S_{\text{on-shell}}^{\text{ren}}|_{\chi=0}.
\end{equation}
To better understand, we calculate the on-shell actions at the respective boundaries. We find:
\begin{align}
& S_{\text{on-shell}}^{\text{ren}}|_{\chi=0}=\int_{\omega,k}\left[ \frac{3\sqrt{6} i}{b} \left( C^E_{\text{in}}+C^E_{\text{out}} \right)\left( C^O_{\text{in}}+C^O_{\text{out}} \right)\a Q k+\frac{1}{8} \left\{ \left( C^E_{\text{in}}+C^E_{\text{out}} \right)^2 \right. \right.\nn \\
&\qquad\qquad\qquad\qquad \left.    -\left( C^O_{\text{in}}+C^O_{\text{out}} \right)^2  \right\} \left\{ 32 G_1 \omega^2 +\left(16 G_2-27\a^2 \ln 2 \right)k^2 \right\} \nn \\
& \qquad\qquad\qquad\qquad \left. \frac{}{}  +\mathcal{O}\left(\omega^3, \omega^2 k, \omega k^2, k^3 \right)\right] \\
& S_{\text{on-shell}}^{\text{ren}}|_{\chi=1}=\int_{\omega,k}\left[ \frac{3\sqrt{6} i}{b} \left( C^E_{\text{in}}+C^E_{\text{out}} \right)\left( C^O_{\text{in}}+C^O_{\text{out}} \right)\a Q k+\frac{1}{8} \left\{ \left( C^E_{\text{in}}+C^E_{\text{out}} \right)^2 \right. \right.\nn \\
&\qquad\qquad\qquad\qquad \left.    -\left( C^O_{\text{in}}+C^O_{\text{out}} \right)^2  \right\} \left\{ 32 G_1 \omega^2 +\left(16 G_2-27\a^2 \ln 2 \right)k^2 \right\} \nn \\
& \qquad\qquad\qquad\qquad \left.  \frac{}{}+ \frac{4i}{b} \left( C^E_{\text{in}}C^E_{\text{out}}-C^O_{\text{in}}C^O_{\text{out}} \right) \b\omega^2+\mathcal{O}\left(\omega^3, \omega^2 k, \omega k^2, k^3 \right)\right].
\end{align}
From here, we can obtain
\begin{equation}
S_{\text{on-shell}}^{\text{SK}}=\int_{\omega,k} \left[ \frac{4i}{b} \left( C^E_{\text{in}}C^E_{\text{out}}-C^O_{\text{in}}C^O_{\text{out}} \right) \b\omega^2+\mathcal{O}\left(\omega^3, \omega^2 k, \omega k^2, k^3 \right)\right].
\end{equation}
We can express this in terms of left/right basis by using Eq. \eqref{LR}. We find
\begin{align}
& S_{\text{on-shell}}^{\text{SK}} =\int_{\omega,k}  \left[ -\frac{4i e^{\b\omega}}{b \left(e^{\b\omega}-1 \right)^2}\left\{ C^E_L \left( C^E_L-C^E_R \right)-C^O_L \left( C^O_L-C^O_R \right)+\left[ C^O_R \left( C^O_L-C^O_R \right) \right. \right.\right. \nn \\
&\qquad\qquad\qquad \left. \left. \left. -C^E_R \left( C^E_L-C^E_R \right)e^{\b\omega} \right] \b\omega^2\right\}+\frac{}{}\mathcal{O}\left(\omega^3, \omega^2 k, \omega k^2, k^3 \right) \right] \label{SKLR}
\end{align}
We can also express this in terms of Keldysh (average/difference) basis
\begin{equation}
C^{E,O}_{A}=\frac{1}{2}\left( C^{E,O}_{L}+C^{E,O}_{R} \right), \ \ C^{E,O}_{D}=\left( C^{E,O}_{L}-C^{E,O}_{R} \right). 
\end{equation}
In terms of this basis we can write Eq. \eqref{SKLR} as
\begin{align}
& S_{\text{on-shell}}^{\text{SK}} =\int_{\omega,k} \left[ -\frac{2i e^{\b\omega}}{b \left(e^{\beta\omega}-1 \right)^2} \left\{ C^E_D \left( 2C^E_A+C^E_D \right)-C^O_D \left(2 C^O_A+C^O_D \right)+\left[C^O_D \left( 2 C^O_A-C^O_D\right) \right. \right. \right. \nn \\ 
&\left. \left. \left. -C^E_D \left( 2 C^E_A-C^E_D \right) \right]  e^{\beta\omega} \right\} \b\omega^2  +\frac{}{}\mathcal{O}\left(\omega^3, \omega^2 k, \omega k^2, k^3 \right) \right].
\end{align}
This vanishes when we set $C^E_D=C^O_D=0$ as expected for a Keldysh functional. The important piece of information we can extract is that contribution from the Chern-Simons term gets canceled. Hence, the Schwinger-Keldysh functional is insensitive to the anomaly, at least up to the order that we considered here. Although we have not computed the higher order terms in derivative expansion, we believe the cancellation of anomaly will also persist for the higher order. This statement, however, requires careful scrutiny checks which we keep as a future work. 
\section{Conclusion}
In this work we have studied the Maxwell-Chern-Simons theory in a $5$-dimensional space-time: a charged black-brane which asymptotes to a AdS$_5$ near $r=\infty$ which is the boundary. The gauge field only has time component. Via the AdS/CFT correspondence, we are considering a current in the dual field theory side with a non-trivial chemical potential which is the difference between the gauge field value evaluated at the boundary and horizon. \\
We have considered perturbation of the gauge field. This perturbation can be classified as vector and scalar modes depending on their transformation properties with respect to the spatial $\mathbb{R}^3$ metric. In Appendix  \ref{scalar}, we have shown that the Chern-Simons term does not contribute for scalar perturbation. Therefore, the non-trivial perturbation comes from the vector sector only. This is  further classified as even or odd depending on their transformation under parity.   \\
We have found the equations of motion for the vector fluctuations. These equations are different for the even and odd sector. Unlike the Maxwell sector, the presence of the Chern-Simons term mixes this even and odd perturbation. To solve these equations, we have performed a derivative expansion which we call the hydrodynamic expansion. The details of this procedure is presented in Appendix \ref{details}. While in the first order we have found an exact form of the solution, in the second order the integrations become intractable. For this case, we have performed an asymptotic expansion of the integrand, and then we integrate. This procedure enables us to find the asymptotic expansion of the formal solution. \\
We have found a quadratic effective action \eqref{2dev} variation of which gives the vector equations of motion. This come from the Maxwell-Chern-Simons action, and then expanding it to the quadratic order. While the Maxwell part of the action is parity preserving, the Chern-Simons part mixes parity. Later, we show that the expectation value of the divergence of the boundary current is non-zero, and therefore the current is anomalous. We interpret the time derivative as the induced ``electric field", and the spatial derivative to be induced ``magnetic field". Therefore the anomaly is reminiscent of ABJ anomaly of quantum field theory. \\
We have computed the on-shell action which via the AdS/CFT correspondence is the generating functional of the dual current. We have found that parity preserving part comes from Maxwell term. The Chern-Simons term mixes parity, and gives rises to the even-odd correlator. \\
The bulk geometry can be complexified to find the Schwinger-Keldysh functional. We have used a complex radial contour, with the end point boundaries corresponding to the left and right sector of the Schwinger-Keldysh functional. Normal to these boundaries are oppositely oriented, and after the appropriate subtraction of the on-shell actions, we have obtained the Schwinger-Keldysh functional. In the Keldysh basis (average/difference basis), we have found that the Schwinger-Keldysh functional becomes zero when we set the difference source to be zero. This is an expected feature of Keldysh functional. Upon the subtraction of the on-shell actions, we found that the Chern-Simons term gets canceled. This implies that the Schwinger-Keldysh function is evanescent to the anomalous term up to the order we have considered, and we expect that this persists for the higher order terms as well.   \\
There are several interesting arenas to explore as future works. In this work, we have not considered the backreaction of the gauge field on the metric. The effects of this gauge field on metric is an interesting problem that we want to study in near future. Depending on the parity of the gauge field, there should be two different dispersion relations. It would be intereseting to study these questions together with the Schwinger-Keldysh analysis. We wish to report the answers to this question in near future. \\
Another study that we wish to pursue is how to address the question of vacuum decay or phase transition in the context of Schwinger-Keldysh physics. Some of the work have been done, and presented in \cite{calzetta_hu_2008, kamenev_2011}, but the effect of strong coupling is not properly understood. Here holography comes into the role, and numerous works have been done under the name Coleman-de Luccia (CdL) tunneling (see \cite{Ghosh:2021lua} for a recent work). It would be interesting to understand the vacuum decay via CdL instanton in the Schwinger-Keldysh formalism. We hope to report this in a future work. 
\section*{Acknowledgments} 
We would like to thank Karl Landsteiner for clarifying issues with the  holographic renormalization to us. We also express our gratitude to R. Loganayagam for illuminating discussions. We also thank Irfan Sadat Ameen  for checking some of the calculations at initial stages of this work. \\
JKG dedicates this paper to his beloved mother Mrs. Shitali Ghosh for her unconditional love, and lifelong support. 
\appendix
 \section{Coordinate Transformations} \label{Coord}
In this appendix, we will find a coordinate transformation between radial and Eddington-Finkelstein coordinates. First we write the AdS-black-brane metric in radial coordinates as follows:
\begin{align}
ds^2&=\frac{d\r^2}{h(\r)a^2(\r)}+a^2(\r)\left[-h(\r)dt^2+d\vec{x}^2 \right] \nonumber \\
&=a^2(\r) h(\r) \left[ -dt^2+\frac{d\r^2}{a^4(\r) h^2(\r)} \right]+a^2(\r)d\vec{x}^2 \ . 
\end{align}
The tortoise coordinate is defined as 
\begin{equation}
\frac{d\r}{a^2(\r)h(\r)}=dr_{\star} \ . 
\end{equation}
In terms of this coordinate, the above metric can be written as 
\begin{equation}
ds^2=a^2(\r) h(\r)\left(-dt+dr_{\star} \right) \left(dt+dr_{\star} \right)+a^2(\r)d\vec{x}^2 \ . 
\end{equation}
Now we define
\begin{equation}
dt+dr_{\star}=dv , \quad d\r=dr \ . 
\end{equation}
In terms of these coordinates, the above metric can be written as
\begin{equation}
ds^2=2dv \ dr -a^2(r)h(r) dv^2+a^2(r)d\vec{x}^2 \ . 
\end{equation}
Setting $a(r)=r$ and $h(r)=f(r)$, we can retrieve the metric \eqref{metans} that we started with. 
\section{Details of  finding the  solutions in derivative expansion}\label{details}
In this appendix we will present the details of the calculation to find the solutions in the derivative expansion. Two equations that we intend to solve are Eqs. \eqref{Vec1} and \eqref{Vec2}. We solve these in a derivative expansion as a double series in $\omega$ and $k$ given by Eqs. \eqref{expE}-\eqref{expO}. \\
\textbf{Zeroth order}\\
For the zeroth order, there is no source term present in the right hand side of Eqs. \eqref{GenE}-\eqref{GenO}. We also impose regularity at the horizon. Imposing this, we find that 
\begin{equation}
 \Psi^{(O)}_{E,O}=\frac{C_{E,O}}{r}
 \end{equation} 
 where $C_E$ and $C_O$ are the integration constants. \\
 \textbf{Order $\omega$} \\
At order $\omega$, we need to solve the following equations: 
\begin{align}
& \frac{d}{dr}\left[ r^3 f \frac{d}{dr} \left( r \Psi_{E}^{\left( \omega \right)
} \right) \right]=2 ir^{1/2}\frac{d}{dr}\left[ r^{3/2} \Psi_{E}^{(0)} \right], \\
& \frac{d}{dr}\left[ r^3 f \frac{d}{dr} \left( r \Psi_{O}^{\left( \omega \right)
} \right) \right]=2 ir^{1/2}\frac{d}{dr}\left[ r^{3/2} \Psi_{O}^{(0)} \right] . 
\end{align}
Since the zeroth order solution was: 
\begin{equation}
\Psi_{E,O}^{(0)}(r)=\frac{C_{E,O}}{r}, 
\end{equation}
the source is: 
\begin{equation}
J_{E,O}^{(\omega)}(r)= i C_{E,O} . 
\end{equation}
Therefore the order $\omega$ solution is: 
\begin{equation}
\Psi_{E,O}^{(\omega)}(r)=-i b^2\frac{ C_{E,O} }{br} F_1(br)
\end{equation}
where 
\begin{equation}
F_1(br)=\int_{br}^\infty d\eta \frac{\eta-1}{\eta^3 f\left(\frac{\eta}{b} \right)}. 
\end{equation} 
This can be integrated to find: 
\begin{align}
& F_1(br)=\left[\frac{1}{\sqrt{2} \left(Q^2-2\right) Z} \left\{ \left(\frac{\left(Z+1\right) Q^2-Z+1}{\sqrt{1-Z}} \right) \right. \right. \nn \\
& \tan ^{-1}\left(\frac{\sqrt{2} \eta }{\sqrt{1-Z}}\right)+ \left( \frac{\left(Z-1\right) Q^2-Z-1}{\sqrt{Z+1}} \right)+\left. \tan ^{-1}\left(\frac{\sqrt{2} \eta }{\sqrt{Z+1}}\right)\right\} \nn \\
& +\frac{1}{4 \left(Q^2-2\right) Z} \left\{ -Z \log \left[ 4 \left(\eta ^4+\eta ^2-Q^2\right)\right]\left. +\left(1-2 Q^2\right) \log \left(\frac{2 \eta ^2-Z+1}{2 \eta ^2+Z+1}\right) \right\} \right. \nn \\
&\left. +\frac{\log (\eta +1)}{Q^2-2} \right]_{br}^\infty.
\end{align}
Although this looks complicated, the near boundary expansion is quite simple! The near boundary $r\rightarrow \infty$ expansion is: 
\begin{align}
\Psi_{E,O}^{(\omega)}(r)=-i C_{E,O} b^2 \left[\frac{1}{(br)^2}-\frac{1}{2 (br)^3}+\frac{1+Q^2}{5 (br)^6}-\frac{1+Q^2}{6 (br)^7}+\mathcal{O}\left(\frac{1}{(br)^8} \right) \right].
\end{align}
\subsubsection*{Order $k$}
At order $k$, we have these two equations
\begin{align}
& \frac{d}{dr}\left[ r^3 f \frac{d}{dr} \left( r \Psi_{E}^{\left( k \right)
} \right) \right]=\frac{3i \alpha \xi' r}{2} \Psi_O^{(0)}, \\
& \frac{d}{dr}\left[ r^3 f \frac{d}{dr} \left( r \Psi_{O}^{\left( k \right)
} \right) \right]=-\frac{3i \alpha \xi' r}{2} \Psi_E^{(0)}. 
\end{align}
By following the same procedure we can find the solution: 
\begin{equation}
\Psi^{(k)}_{E,O}(r)=\pm \left( \frac{3}{2} \right)^{3/2}i \a C_{O,E} Q b^2 \frac{1}{br} F_2 (br)
\end{equation}
where 
\begin{equation}
F_2 (br)=\int_{br}^{\infty} d\eta \frac{\eta^2 -1}{\eta^5 f\left( \frac{\eta}{b} \right)}. 
\end{equation}
This can be integrated to find: 
\begin{align}
F_2 (br)& =-\frac{1}{Z} \left[\tanh^{-1}\left(\frac{1+2\eta^2}{Z} \right) \right]_{br}^{\infty} \nn \\
&= \frac{1}{Z}\left\{\frac{i\pi}{2}+  \tanh^{-1}\left(\frac{1+2(br)^2}{Z} \right) \right\}.
\end{align}
Therefore
\begin{equation}
\Psi^{(k)}_{E,O}(r)=\pm \left( \frac{3}{2} \right)^{3/2} \frac{i \a C_{O,E} Q b^2 }{Z} \frac{1}{br }  \left\{\frac{i\pi}{2}+  \tanh^{-1}\left(\frac{1+2(br)^2}{Z} \right) \right\} .
\end{equation}
The near boundary $r\rightarrow\infty$ expansion is
\begin{align}
\Psi^{(k)}_{E,O}(r)=\pm \left( \frac{3}{2} \right)^{3/2} i \a C_{O,E} Q b^2 \left[ \frac{1}{2(br)^3}-\frac{1}{4 (br)^5}+\frac{1+Q^2}{6 (br)^7 }+\mathcal{O}\left(\frac{1}{(br)^8} \right) \right].
\end{align}
\textbf{Quadratic order}\\
Now we will move to find the second order solutions. These will be determined by solving equations \eqref{Vec1}-\eqref{Vec2} where the source term is composed from the first order solutions. For the solutions of second order, we could not find a closed form expression. However, since we need only the near boundary expansion, by using the strategy mentioned in the main text, we can obtain the near boundary expansions. We will elucidate this below.  
\subsubsection*{Order $\omega^2$}
At this order, the source term is
\begin{equation}
J_{E,O}^{(\omega^2)}(r)=i \left( 3 r \Psi_{E,O}^{(\omega)} +2r^2 \frac{d \Psi_{E,O}^{(\omega)}}{dr}\right).
\end{equation}
After some manipulations, this can be written as: 
\begin{equation}
J_{E,O}^{(\omega^2)}(r)= 2 b C_{E,O}\left(br \right)^{1/2} \frac{d}{d(br)}\left\{ \left(br \right)^{1/2} F_1(br)  \right\} .
\end{equation}
Then applying the general formula \eqref{gen} we can write: 
\begin{equation}
\Psi_{E,O}^{(\omega^2)}=-2 b^3 C_{E,O}\frac{1}{(br)} \int_{br}^{\infty} d\eta \frac{1}{\eta^3 f\left( \eta /b\right)} \int_1^\eta d\kappa \  \sqrt{\kappa}\frac{d}{d\kappa}\left\{ \sqrt{\kappa} F_1(\kappa) \right\}
\end{equation} 
Near the boundary $r\rightarrow \infty $, the function behaves as:
\begin{align}
\Psi_{E,O}^{(\omega^2)}=\frac{b^3}{4} C_{E,O}\left\{\frac{4 G_1-1}{(br)^3}+\frac{2 \ln (br)}{(br)^3} \right\}+\frac{ b^3}{2}C_{E,O} \frac{1}{(br)^4} \nonumber \\
+\frac{(1+Q^2) b^3}{360}C_{E,O}\left\{ \frac{120 G_1-77}{(br)^7}+\frac{60 \ln (br)}{(br)^7} \right\}+\mathcal{O}\left(\frac{1}{(br)^8} \right)
\end{align}
where $G_1$ is the value of the integral $\int d\kappa \sqrt{\kappa}\frac{d}{d\kappa}\left\{ \sqrt{\kappa} F_1(\kappa) \right\}  $ evaluated at $\kappa=1$. This is given by
\begin{align}
G_1& =-\frac{1}{4 Z} \coth^{-1}\left( \frac{3}{Z} \right)-\frac{1}{16 Q \left(Q^2-2\right)} \left[ \sqrt{2} \pi \left\{ -i \sqrt{\frac{1}{Z}+\frac{1}{Z^2}}+\sqrt{\frac{Z-1}{Z^2}} \right. \right. \nn \\
& +\sqrt{Z-1}+i \sqrt{Z+1}-Q^2\left( i \sqrt{\frac{1}{Z}+\frac{1}{Z^2}}+\sqrt{Z-1}+i \sqrt{Z+1} \right. \nn \\
& \left. \left. -\sqrt{\frac{Z-1}{Z^2}} \right) \right\} +Q \ln 16- \left. \frac{}{}+2 Q \left(Q^2-2\right) \left(\log \left(3-Z\right)+\log \left(Z+3\right)\right)  \right] .\label{G1}
\end{align}
\subsubsection*{Order $k^2$}
The source term at this order is given by: 
\begin{equation}
J_{E,O}^{(k^2)}=\Psi_{E,O}^{(0)}\pm \frac{3i\a \xi' r}{2} \Psi_{O,E}^{(k)}. 
\end{equation}
Using different expressions, we can write the this source as: 
\begin{equation}
J_{E,O}^{(k^2)}= C_{E,O} b \left\{\frac{1}{br}-\frac{27\a^2 Q^2}{4 (br)^3}F_2 (br) \right\}.
\end{equation}
Inserting this into the general expression for the solution \eqref{gen} we can write: 
\begin{equation}
\Psi_{E,O}^{(k^2)} =- C_{E,O} b^3 \frac{1}{br} \int_{br}^\infty d\eta \frac{1}{\eta^3 f\left(\frac{\eta}{b} \right)} \int_1^\eta d\kappa  \left\{\frac{1}{\kappa}-\frac{27\a^2 Q^2}{4 \kappa^3}F_2 (\kappa) \right\}
 \end{equation}
 This can be simplified a bit further
 \begin{equation}
 \Psi_{E,O}^{(k^2)} =- C_{E,O} b^3 \frac{1}{br}\int_{br}^\infty d\eta \frac{\ln\eta}{\eta^3 f\left(\frac{\eta}{b} \right)}+\frac{27 C_{E,O}b^3 \a^2 Q^2 }{4 br}\int_{br}^\infty d\eta \frac{1}{\eta^3 f\left(\frac{\eta}{b} \right)}\int_1^\eta d\kappa \frac{F_2 (\kappa)}{\kappa^3}.
 \end{equation}
 Using the techniques outlined in the main text we can extract the near boundary $r\rightarrow\infty$ expansion 
 \begin{align}
 \Psi_{E,O}^{(k^2)} = & C_{E,O} \frac{-27 \alpha ^2 \ln 2-16 \ln (b r)+16 G_2-8}{32 r^3}+\frac{C_{E,O}}{576 b^4 r^7}\left\{96 G_2 \left(Q^2+1\right)-16 \left(Q^2+1\right) \right. \nn \\
& \left. -81 \alpha ^2 \left[ Q^2 (1+\ln 4)+\ln 4\right]-96 \left(Q^2+1\right) \ln (b r) \right\}+\mathcal{O}\left(\frac{1}{(br)^8} \right)
 \end{align}
 where $G_2$ is the value of $\int d\kappa \left\{\frac{1}{\kappa}-\frac{27\a^2 Q^2}{4 \kappa^3}F_2 (\kappa) \right\} $ evaluated at $\kappa=1$. Explicitly, this is given by
 \begin{align}
 G_2& =\frac{27 \alpha ^2}{32 Z}\left[ 2 i \pi  Q^2+4 Q^2 \tanh ^{-1}\left(\frac{3}{Z}\right)-2 \coth ^{-1}\left(\frac{3}{Z}\right)  \right. \nn \\
 & \left. Z \left\{ \ln \left(3-Z\right)+\ln \left(Z+3\right)\right\} \right]. \label{G2}
 \end{align}
 \subsubsection*{Order $\omega k$}
 At this order the source term is 
 \begin{equation}
 J_{E,O}^{(\omega k)}=3ir \Psi_{E,O}^{(k)}+2ir^2 \frac{d \Psi_{E,O}^{(k)}}{dr}\pm \frac{3i\a \xi' r}{2} \Psi_{O,E}^{(\omega)}. 
 \end{equation}
 Using different expressions above and after some manipulations we can write: 
 \begin{equation}
 J_{E,O}^{(\omega k)}= \mp 2 \left(\frac{3}{2} \right)^{3/2} C_{O,E} \a Q b \left[  (br)^{\frac{1}{2}} \frac{d}{d(br)} \left\{ (br)^{\frac{1}{2}} F_2 (br) \right\}+ \frac{F_1(br)}{(br)^3}  \right]. 
 \end{equation}
 Then we can write from the general solution \eqref{gen}
 \begin{equation}
 \Psi_{E,O}^{(\omega k)}=\pm 2 \left(\frac{3}{2} \right)^{\frac{3}{2}} C_{O,E} \a Q b^3 \frac{1}{br} \int_{br}^\infty d\eta \frac{1}{\eta^3 f\left(\frac{\eta}{b} \right)} \int_1^\eta d\kappa  \left[  \sqrt{\kappa} \frac{d}{d\kappa} \left\{ \sqrt{\kappa} F_2 (\kappa) \right\}+ \frac{F_1(\kappa)}{\kappa^3}  \right].
 \end{equation}
 Again, using the method outlined before we can extract the near boundary asymptotics: 
 \begin{align}
 \Psi_{E,O}^{(\omega k)}=\pm b^3 \left( \frac{3}{2}\right)^{\frac{1}{2}} \alpha C_{O,E} \left\{ \frac{f_3 (Q)}{(br)^3}+\frac{3Q}{4 (br)^4}-\frac{3Q}{8 (br)^6} \right\}+\mathcal{O}\left(\frac{1}{r^7} \right)
 \end{align}
 where: 
 \begin{align}
 & f_3(Q)=-\frac{3}{32 \left(Q^2-2 \right) Z^2  } \left[ 16 G_3 Q \left(Q^2-2\right) Z^2+\sqrt{2}\pi \left\{ -\sqrt{-1+Z} \right.  \right. \nn \\
 & -\sqrt{ Z^2  \left(-1+Z\right)}+Q \left(i\sqrt{-1+Z}-i\sqrt{Z^2( -1+Z) }  -\sqrt{1+Z} \right. \nn \\
 & +\sqrt{Z^2(1+Z)}+2 Q \left(-2 \sqrt{-1+Z}+Q \left(2i\sqrt{-1+Z}+i \sqrt{Z^2(-1+Z) } \right. \right. \nn \\
 &\left. \left. \left. \left. \left. -2 \sqrt{1+Z} +\sqrt{Z^2(1+Z) } \right) \right) \right)\right\}-4 Q Z^2 \ln 2  \right] \label{f3}
 \end{align}
 and $G_3$ is the value of the integral $\int d\kappa \left[ 3 \sqrt{\kappa} \frac{d}{d\kappa} \left\{ \sqrt{\kappa} F_2 (\kappa) \right\}+9 \frac{F_1(\kappa)}{\kappa^3}  \right] $ evaluated at $\kappa=1$. The explicit expression of $G_3$ can be found by straight-forward calculation. But this is long, and not important for the result. So we refrain to record this here. 
\section{Scalar sector} \label{scalar}
 In this appendix we will study the scalar sector. First we write the perturbation as: 
 \begin{align}
 & a_r=\int \frac{d\omega}{2\pi}\frac{d^3k}{(2\pi)^3} \Phi_r (r,\omega,k)\ S\left(\vec{k},\vec{x} \right), \\
&  a_v=\int \frac{d\omega}{2\pi}\frac{d^3k}{(2\pi)^3} \Phi_t (r,\omega,k)\ S\left(\vec{k},\vec{x} \right), \\
& a_i = r \int \frac{d\omega}{2\pi}\frac{d^3k}{(2\pi)^3} \Phi_S (r,\omega,k)\ \p_i S\left(\vec{k},\vec{x} \right).
 \end{align}
 Here $S(\vec{k},\vec{x})$ is a scalar harmonic with the following property: 
 \begin{equation}
 \left( \nabla^2+k^2 \right)S=0.
 \end{equation}
 The field-strength tensor is given by: 
 \begin{align}
 F& =2\left\{-\xi'+\left(i \omega \Phi_r+\Phi_t'  \right)e^{-i \omega v} S \right\}dr\wedge dv+2 \left(-\Phi_r+\Phi_S+r \Phi_S' \right)e^{-i \omega v}\p_i S\   dr\wedge dx^i \nonumber \\
&\qquad\qquad\qquad\qquad\qquad\qquad\qquad\qquad  -2\left(i \omega r \Phi_s+ \Phi_t \right)e^{-i \omega v} \p_i S \ dv\wedge dx^i.
 \end{align}
 From this expression this is apparent that 
 \begin{equation}
  F\wedge F=0,
  \end{equation} 
  and consequently the Chern-Simons term does not contribute at all in the scalar sector. The equations for the fluctuation is therefore:  
  \begin{equation}
  \p_a \left(\sqrt{-g}F^{ab} \right)=0
  \end{equation}
  where we take the linear order part. This has been analyzed in \cite{Ghosh:2020lel}.
 
\bibliography{ref.bib}

\providecommand{\href}[2]{#2}\begingroup\raggedright\begin{thebibliography}{10}

\bibitem{Maldacena:1997re}
J.~M. Maldacena, \emph{{The Large N limit of superconformal field theories and
  supergravity}}, \href{https://doi.org/10.1023/A:1026654312961}{\emph{Adv.
  Theor. Math. Phys.} {\bfseries 2} (1998) 231}
  [\href{https://arxiv.org/abs/hep-th/9711200}{{\ttfamily hep-th/9711200}}].

\bibitem{Aharony:1999ti}
O.~Aharony, S.~S. Gubser, J.~M. Maldacena, H.~Ooguri and Y.~Oz, \emph{{Large N
  field theories, string theory and gravity}},
  \href{https://doi.org/10.1016/S0370-1573(99)00083-6}{\emph{Phys. Rept.}
  {\bfseries 323} (2000) 183}
  [\href{https://arxiv.org/abs/hep-th/9905111}{{\ttfamily hep-th/9905111}}].

\bibitem{Casalderrey-Solana:2011dxg}
J.~Casalderrey-Solana, H.~Liu, D.~Mateos, K.~Rajagopal and U.~A. Wiedemann,
  \emph{{Gauge/String Duality, Hot QCD and Heavy Ion Collisions}}. Cambridge
  University Press, 2014,
  \href{https://doi.org/10.1017/CBO9781139136747}{10.1017/CBO9781139136747},
  [\href{https://arxiv.org/abs/1101.0618}{{\ttfamily 1101.0618}}].

\bibitem{Gubser:1998bc}
S.~S. Gubser, I.~R. Klebanov and A.~M. Polyakov, \emph{{Gauge theory
  correlators from noncritical string theory}},
  \href{https://doi.org/10.1016/S0370-2693(98)00377-3}{\emph{Phys. Lett. B}
  {\bfseries 428} (1998) 105}
  [\href{https://arxiv.org/abs/hep-th/9802109}{{\ttfamily hep-th/9802109}}].

\bibitem{Witten:1998qj}
E.~Witten, \emph{{Anti-de Sitter space and holography}},
  \href{https://doi.org/10.4310/ATMP.1998.v2.n2.a2}{\emph{Adv. Theor. Math.
  Phys.} {\bfseries 2} (1998) 253}
  [\href{https://arxiv.org/abs/hep-th/9802150}{{\ttfamily hep-th/9802150}}].

\bibitem{calzetta_hu_2008}
E.~A. Calzetta and B.-L.~B. Hu, \emph{Nonequilibrium Quantum Field Theory},
  Cambridge Monographs on Mathematical Physics. Cambridge University Press,
  2008,
  \href{https://doi.org/10.1017/CBO9780511535123}{10.1017/CBO9780511535123}.

\bibitem{kamenev_2011}
A.~Kamenev, \emph{Field Theory of Non-Equilibrium Systems}. Cambridge
  University Press, 2011,
  \href{https://doi.org/10.1017/CBO9781139003667}{10.1017/CBO9781139003667}.

\bibitem{Herzog:2002pc}
C.~P. Herzog and D.~T. Son, \emph{{Schwinger-Keldysh propagators from AdS/CFT
  correspondence}},
  \href{https://doi.org/10.1088/1126-6708/2003/03/046}{\emph{JHEP} {\bfseries
  03} (2003) 046} [\href{https://arxiv.org/abs/hep-th/0212072}{{\ttfamily
  hep-th/0212072}}].

\bibitem{Glorioso:2018mmw}
P.~Glorioso, M.~Crossley and H.~Liu, \emph{{A prescription for holographic
  Schwinger-Keldysh contour in non-equilibrium systems}},
  \href{https://arxiv.org/abs/1812.08785}{{\ttfamily 1812.08785}}.

\bibitem{Chakrabarty:2019aeu}
B.~Chakrabarty, J.~Chakravarty, S.~Chaudhuri, C.~Jana, R.~Loganayagam and
  A.~Sivakumar, \emph{{Nonlinear Langevin dynamics via holography}},
  \href{https://doi.org/10.1007/JHEP01(2020)165}{\emph{JHEP} {\bfseries 01}
  (2020) 165} [\href{https://arxiv.org/abs/1906.07762}{{\ttfamily
  1906.07762}}].

\bibitem{Loganayagam:2020eue}
R.~Loganayagam, K.~Ray and A.~Sivakumar, \emph{{Fermionic Open EFT from
  Holography}},  \href{https://arxiv.org/abs/2011.07039}{{\ttfamily
  2011.07039}}.

\bibitem{Jana:2020vyx}
C.~Jana, R.~Loganayagam and M.~Rangamani, \emph{{Open quantum systems and
  Schwinger-Keldysh holograms}},
  \href{https://doi.org/10.1007/JHEP07(2020)242}{\emph{JHEP} {\bfseries 07}
  (2020) 242} [\href{https://arxiv.org/abs/2004.02888}{{\ttfamily
  2004.02888}}].

\bibitem{Loganayagam:2020iol}
R.~Loganayagam, K.~Ray, S.~K. Sharma and A.~Sivakumar, \emph{{Holographic KMS
  relations at finite density}},
  \href{https://doi.org/10.1007/JHEP03(2021)233}{\emph{JHEP} {\bfseries 03}
  (2021) 233} [\href{https://arxiv.org/abs/2011.08173}{{\ttfamily
  2011.08173}}].

\bibitem{Ghosh:2020lel}
J.~K. Ghosh, R.~Loganayagam, S.~G. Prabhu, M.~Rangamani, A.~Sivakumar and
  V.~Vishal, \emph{{Effective field theory of stochastic diffusion from
  gravity}}, \href{https://doi.org/10.1007/JHEP05(2021)130}{\emph{JHEP}
  {\bfseries 05} (2021) 130}
  [\href{https://arxiv.org/abs/2012.03999}{{\ttfamily 2012.03999}}].

\bibitem{He:2021jna}
T.~He, R.~Loganayagam, M.~Rangamani and J.~Virrueta, \emph{{An effective
  description of momentum diffusion in a charged plasma from holography}},
  \href{https://doi.org/10.1007/JHEP01(2022)145}{\emph{JHEP} {\bfseries 01}
  (2022) 145} [\href{https://arxiv.org/abs/2108.03244}{{\ttfamily
  2108.03244}}].

\bibitem{He:2022jnc}
T.~He, R.~Loganayagam, M.~Rangamani, A.~Sivakumar and J.~Virrueta, \emph{{The
  timbre of Hawking gravitons: an effective description of energy transport
  from holography}},  \href{https://arxiv.org/abs/2202.04079}{{\ttfamily
  2202.04079}}.

\bibitem{He:2022deg}
T.~He, R.~Loganayagam, M.~Rangamani and J.~Virrueta, \emph{{An effective
  description of charge diffusion and energy transport in a charged plasma from
  holography}},  \href{https://arxiv.org/abs/2205.03415}{{\ttfamily
  2205.03415}}.

\bibitem{Bu:2020jfo}
Y.~Bu, T.~Demircik and M.~Lublinsky, \emph{{All order effective action for
  charge diffusion from Schwinger-Keldysh holography}},
  \href{https://doi.org/10.1007/JHEP05(2021)187}{\emph{JHEP} {\bfseries 05}
  (2021) 187} [\href{https://arxiv.org/abs/2012.08362}{{\ttfamily
  2012.08362}}].

\bibitem{Bu:2021jlp}
Y.~Bu and B.~Zhang, \emph{{Schwinger-Keldysh effective action for a
  relativistic Brownian particle in the AdS/CFT correspondence}},
  \href{https://doi.org/10.1103/PhysRevD.104.086002}{\emph{Phys. Rev. D}
  {\bfseries 104} (2021) 086002}
  [\href{https://arxiv.org/abs/2108.10060}{{\ttfamily 2108.10060}}].

\bibitem{Bu:2022esd}
Y.~Bu, X.~Sun and B.~Zhang, \emph{{Holographic Schwinger-Keldysh field theory
  of SU(2) diffusion}},  \href{https://arxiv.org/abs/2205.00195}{{\ttfamily
  2205.00195}}.

\bibitem{Banerjee:2022aub}
A.~Banerjee, T.~Mitra and A.~Mukhopadhyay, \emph{{Correlation functions of the
  Bjorken flow in the holographic Schwinger-Keldysh approach}},
  \href{https://arxiv.org/abs/2207.00013}{{\ttfamily 2207.00013}}.

\bibitem{Bertlmann:1996xk}
R.~A. Bertlmann, \emph{{Anomalies in quantum field theory}}. 1996.

\bibitem{Fujikawa:2004cx}
K.~Fujikawa and H.~Suzuki, \emph{{Path integrals and quantum anomalies}}. 2004,
  \href{https://doi.org/10.1093/acprof:oso/9780198529132.001.0001}{10.1093/acprof:oso/9780198529132.001.0001}.

\bibitem{Fukushima:2008xe}
K.~Fukushima, D.~E. Kharzeev and H.~J. Warringa, \emph{{The Chiral Magnetic
  Effect}}, \href{https://doi.org/10.1103/PhysRevD.78.074033}{\emph{Phys. Rev.
  D} {\bfseries 78} (2008) 074033}
  [\href{https://arxiv.org/abs/0808.3382}{{\ttfamily 0808.3382}}].

\bibitem{Kharzeev:2013ffa}
D.~E. Kharzeev, \emph{{The Chiral Magnetic Effect and Anomaly-Induced
  Transport}}, \href{https://doi.org/10.1016/j.ppnp.2014.01.002}{\emph{Prog.
  Part. Nucl. Phys.} {\bfseries 75} (2014) 133}
  [\href{https://arxiv.org/abs/1312.3348}{{\ttfamily 1312.3348}}].

\bibitem{Landsteiner:2016led}
K.~Landsteiner, \emph{{Notes on Anomaly Induced Transport}},
  \href{https://doi.org/10.5506/APhysPolB.47.2617}{\emph{Acta Phys. Polon. B}
  {\bfseries 47} (2016) 2617}
  [\href{https://arxiv.org/abs/1610.04413}{{\ttfamily 1610.04413}}].

\bibitem{Kharzeev:2020jxw}
D.~E. Kharzeev and J.~Liao, \emph{{Chiral magnetic effect reveals the topology
  of gauge fields in heavy-ion collisions}},
  \href{https://doi.org/10.1038/s42254-020-00254-6}{\emph{Nature Rev. Phys.}
  {\bfseries 3} (2021) 55} [\href{https://arxiv.org/abs/2102.06623}{{\ttfamily
  2102.06623}}].

\bibitem{Ghosh:2021naw}
J.~K. Ghosh, S.~Grieninger, K.~Landsteiner and S.~Morales-Tejera, \emph{{Is the
  chiral magnetic effect fast enough?}},
  \href{https://doi.org/10.1103/PhysRevD.104.046009}{\emph{Phys. Rev. D}
  {\bfseries 104} (2021) 046009}
  [\href{https://arxiv.org/abs/2105.05855}{{\ttfamily 2105.05855}}].

\bibitem{Policastro:2001yc}
G.~Policastro, D.~T. Son and A.~O. Starinets, \emph{{The Shear viscosity of
  strongly coupled N=4 supersymmetric Yang-Mills plasma}},
  \href{https://doi.org/10.1103/PhysRevLett.87.081601}{\emph{Phys. Rev. Lett.}
  {\bfseries 87} (2001) 081601}
  [\href{https://arxiv.org/abs/hep-th/0104066}{{\ttfamily hep-th/0104066}}].

\bibitem{Policastro:2002se}
G.~Policastro, D.~T. Son and A.~O. Starinets, \emph{{From AdS / CFT
  correspondence to hydrodynamics}},
  \href{https://doi.org/10.1088/1126-6708/2002/09/043}{\emph{JHEP} {\bfseries
  09} (2002) 043} [\href{https://arxiv.org/abs/hep-th/0205052}{{\ttfamily
  hep-th/0205052}}].

\bibitem{Kovtun:2003wp}
P.~Kovtun, D.~T. Son and A.~O. Starinets, \emph{{Holography and hydrodynamics:
  Diffusion on stretched horizons}},
  \href{https://doi.org/10.1088/1126-6708/2003/10/064}{\emph{JHEP} {\bfseries
  10} (2003) 064} [\href{https://arxiv.org/abs/hep-th/0309213}{{\ttfamily
  hep-th/0309213}}].

\bibitem{Banerjee:2008th}
N.~Banerjee, J.~Bhattacharya, S.~Bhattacharyya, S.~Dutta, R.~Loganayagam and
  P.~Surowka, \emph{{Hydrodynamics from charged black branes}},
  \href{https://doi.org/10.1007/JHEP01(2011)094}{\emph{JHEP} {\bfseries 01}
  (2011) 094} [\href{https://arxiv.org/abs/0809.2596}{{\ttfamily 0809.2596}}].

\bibitem{Gubser:2007nd}
S.~S. Gubser and S.~S. Pufu, \emph{{Master field treatment of metric
  perturbations sourced by the trailing string}},
  \href{https://doi.org/10.1016/j.nuclphysb.2007.08.015}{\emph{Nucl. Phys. B}
  {\bfseries 790} (2008) 42}
  [\href{https://arxiv.org/abs/hep-th/0703090}{{\ttfamily hep-th/0703090}}].

\bibitem{Kodama:2003kk}
H.~Kodama and A.~Ishibashi, \emph{{Master equations for perturbations of
  generalized static black holes with charge in higher dimensions}},
  \href{https://doi.org/10.1143/PTP.111.29}{\emph{Prog. Theor. Phys.}
  {\bfseries 111} (2004) 29}
  [\href{https://arxiv.org/abs/hep-th/0308128}{{\ttfamily hep-th/0308128}}].

\bibitem{Landsteiner:2011iq}
K.~Landsteiner, E.~Megias, L.~Melgar and F.~Pena-Benitez, \emph{{Holographic
  Gravitational Anomaly and Chiral Vortical Effect}},
  \href{https://doi.org/10.1007/JHEP09(2011)121}{\emph{JHEP} {\bfseries 09}
  (2011) 121} [\href{https://arxiv.org/abs/1107.0368}{{\ttfamily 1107.0368}}].

\bibitem{Zaanen:2015oix}
J.~Zaanen, Y.-W. Sun, Y.~Liu and K.~Schalm, \emph{{Holographic Duality in
  Condensed Matter Physics}}. Cambridge Univ. Press, 2015.

\bibitem{Ammon:2015wua}
M.~Ammon and J.~Erdmenger, \emph{{Gauge/gravity duality}: {Foundations and
  applications}}. Cambridge University Press, Cambridge, 4, 2015.

\bibitem{Adler:1969gk}
S.~L. Adler, \emph{{Axial vector vertex in spinor electrodynamics}},
  \href{https://doi.org/10.1103/PhysRev.177.2426}{\emph{Phys. Rev.} {\bfseries
  177} (1969) 2426}.

\bibitem{Bell:1969ts}
J.~S. Bell and R.~Jackiw, \emph{{A PCAC puzzle: $\pi^0 \to \gamma \gamma$ in
  the $\sigma$ model}}, \href{https://doi.org/10.1007/BF02823296}{\emph{Nuovo
  Cim. A} {\bfseries 60} (1969) 47}.

\bibitem{Peskin:1995ev}
M.~E. Peskin and D.~V. Schroeder, \emph{{An Introduction to quantum field
  theory}}. Addison-Wesley, Reading, USA, 1995.

\bibitem{Ghosh:2021lua}
J.~K. Ghosh, E.~Kiritsis, F.~Nitti and L.~T. Witkowski, \emph{{Revisiting
  Coleman-de Luccia transitions in the AdS regime using holography}},
  \href{https://doi.org/10.1007/JHEP09(2021)065}{\emph{JHEP} {\bfseries 09}
  (2021) 065} [\href{https://arxiv.org/abs/2102.11881}{{\ttfamily
  2102.11881}}].

\end{thebibliography}\endgroup
\bibliographystyle{JHEP}
\end{document}